\documentclass[]{aa} 
\usepackage{graphicx}
\usepackage{txfonts}
\usepackage{longtable,lscape}
\usepackage[authoryear]{natbib}

\newcommand{\Teff}{\mbox{$T_\mathrm{eff}$}} 
\newcommand{\logg}{\mbox{$\log g$}}

\begin{document}
   \title{High-resolution UVES/VLT spectra of white dwarfs observed
   for the ESO SN Ia Progenitor Survey. III. DA white dwarfs\thanks{Based on 
   data obtained at the Paranal Observatory of
   the European Southern Observatory for programmes 165.H-0588 and
   167.D-0407.}}

\titlerunning{High-resolution spectra observed for the SPY. III. DA white dwarfs}

   \author{D.~Koester
          \inst{1}
	  \and
          B.~Voss
          \inst{2,1}
          \and
          R.~Napiwotzki
          \inst{3}
          \and
          N.~Christlieb\inst{4}
          \and
          D.~Homeier\inst{5}
          \and
          T.~Lisker\inst{6,8}
          \and
          D.~Reimers\inst{7}
          \and
          U.~Heber\inst{8}
          }


   \institute{Institut f\"ur Theoretische Physik und Astrophysik, Universit\"at
              Kiel, 24098 Kiel, Germany
    \and  Zeiss-Planetarium , LWL-Museum f\"ur Naturkunde, 48161 M\"unster, Germany
    \and Centre for Astrophysics Research, University of Hertfordshire,
         College Lane, Hatfield AL10 9AB, UK
    \and Landessternwarte, Zentrum f\"ur Astronomie der Universit\"at Heidelberg, 
         K\"onigstuhl 12, 69117 Heidelberg, Germany
    \and Institut f\"ur Astrophysik, Georg--August--Universit\"at, 
       Friedrich--Hund--Platz 1, 37077 G\"ottingen, Germany 
    \and Astronomisches Rechen--Institut, Zentrum f\"ur Astronomie der 
       Universit\"at Heidelberg, M\"onchhofstr.~12-14, 
       69120 Heidelberg, Germany
    \and Hamburger Sternwarte, Gojenbergsweg 112, 21029 Hamburg, Germany 
    \and Dr.\ Karl Remeis Observatory, University of Erlangen--N\"urnberg, 
       Sternwartstr.~7, 96049 Bamberg, Germany}
   \date{}

   \abstract{The ESO Supernova Ia Progenitor Survey (SPY) took
    high-resolution spectra of more than 1000 white dwarfs and pre-white 
    dwarfs. About two thirds of the stars observed are hydrogen-dominated
    DA white dwarfs. Here we present a catalog and detailed spectroscopic 
    analysis of the DA stars in the SPY.}
   {Atmospheric parameters effective temperature and surface gravity are
   determined for normal DAs. Double-degenerate binaries, DAs with
   magnetic fields or dM companions, are classified and discussed.}
   {The spectra are compared with theoretical model atmospheres using
   a $\chi^2$ fitting technique.}
   {Our final sample contains 615 DAs, which show only hydrogen features
   in their spectra, although some are double-degenerate binaries. 187
   are new detections or classifications. We also find 10 magnetic DAs
   (4 new) and 46 DA+dM pairs (10 new).  }{}

                                                                                  
\maketitle

\section{Introduction}

The ESO SN Ia Progenitor Survey (SPY) is a radial velocity survey that
was conducted to test the double-degenerate channel of the formation
of supernovae Ia. About 800 white dwarfs were observed (most of them
twice) in the course of the survey, assembling a large collection of
high quality white dwarf spectra.  Most of the targets for the SPY
were selected from the White Dwarf catalog \citep{McCook99}, MCS
further on, from the Hamburg/ESO Survey \citep{Wisotzki96,
  Christlieb01}, HES further on, and from the Hamburg Quasar Survey
\citep{Hagen95}, HQS further on. Some additional objects come
  from the Montreal-Cambridge survey \citep{Demers90, Lamontagne00}
  and from the Edinburgh-Cape survey \citep{Kilkenny91}.  A detailed
discussion of the target selection is given in \citet{Napiwotzki01,
  Napiwotzki03}. While some of the input catalogs, e.g. HES and HQS,
have fairly well understood selection criteria, for the combined input
catalog this would be very difficult, since the only criteria used
were ``spectroscopic identification (at least from objective prism
spectra) and B $< 16.5$'' \citep{Napiwotzki01}.

\begin{figure*}[p]
\includegraphics[width=16.6cm]{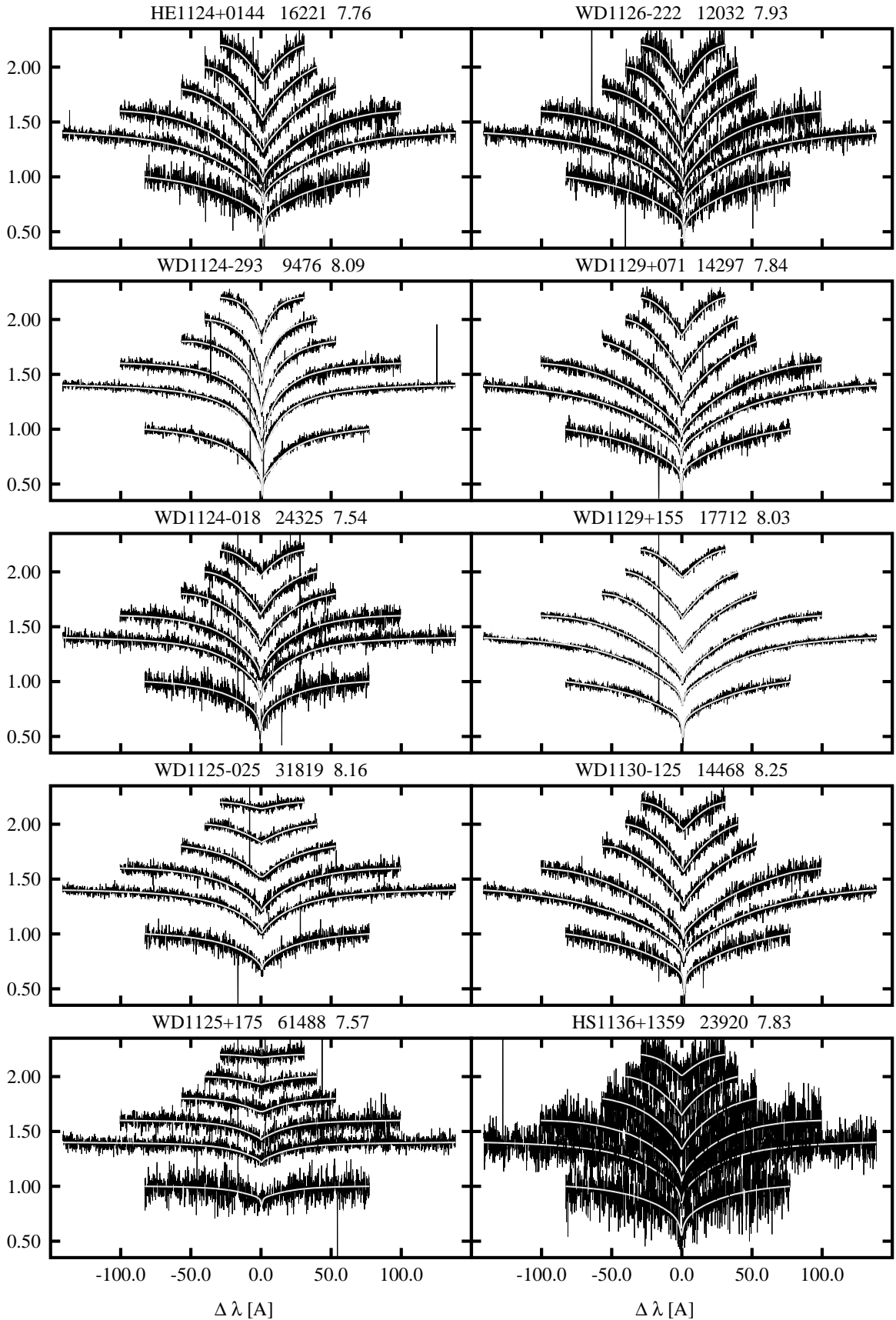}
\caption{Typical example for observations and fit, taking arbitrarily
  the entries 301 to 310 from Table~1. The header of each panel gives
  the name, effective temperature, and surface gravity of the fit.
Shown are the six lowest Balmer lines. Vertical axis is relative
intensity in arbitrary units, higher lines are offset for clarity.
The light grey lines are the models.
  \label{fig1}}
\end{figure*}

The high quality spectra of white dwarfs were employed for a number of
studies beyond the original scope of the SPY, the search for
double-degenerate binaries. \cite{Koester01} derived temperatures
and gravities from a preliminary sample of about 200 objects; 71
helium-rich stars of spectral type DB and DBA were studied in 
\cite{Voss07}, and approximately 60 objects with detected \ion{Ca}{ii}
resonance lines were discussed in \cite{Koester05}. \cite{Pauli03, Pauli06}
studied the 3D kinematics using the SPY data, new DO
and PG1159 stars have been identified by \cite{Werner04}. The 
hot subdwarf population has been studied by \cite{Lisker05} and 
\cite{Stroeer07}.

The number of newly detected white dwarfs is of course dwarfed by the
results from the Sloan Digital Sky Survey \citep[e.g.][]{Kleinman04,
Eisenstein06}.  It is also possible to extract samples with
much better controlled and understood selection effects from the SDSS
data base. Thus, we will not produce yet another white dwarf mass
distribution or luminosity function here. However, our sample contains
much brighter objects than the typical SDSS white dwarf.  These
objects, and in particular the magnetic stars, binaries, or new
variables will be much easier to study in follow-up observations than
the faint SDSS objects. Moreover, the white dwarf parameters presented
here are an important ingredient of the kinematic studies mentioned
above and the analysis and interpretation of the binary results,
starting with the simple distinction between helium core and
carbon/oxygen core white dwarfs.

\section{Observations} 
Since several of the papers above have given a detailed description of
the data properties and further references, we only briefly summarize
this information here. 

The spectra were obtained with UVES, a high resolution echelle
spectrograph at the ESO VLT telescope. UVES was used in a dichroic
mode, resulting in small gaps, $\approx 80$\,\AA~wide, at 4580\,\AA~and
5640\,\AA~in the final merged spectrum. The spectral resolution at
H$\alpha$ is $R=18\,500$ or better, and the $S/N$ per binned pixel
(0.05\,\AA) is $S/N=15$ or higher. The total wavelength range covered
is $\approx$ 3\,500 to 6650~\AA.

The spectra were reduced with the ESO pipeline for UVES, including the
merging of the echelle orders and the wavelength calibration. \cite{Koester01}
found that the quality of these automatically extracted
spectra was very good, except for a quasi-periodic wave-like pattern
that occurs in some of the spectra. The reduction has since been 
further improved by additional processing by collaborators of the SPY
at the University of Erlangen-N\"urnberg. The most important step was
utilizing the featureless spectra of DC white dwarfs to remove almost
completely the large scale variations of the spectral response function
\citep{Napiwotzki01, Napiwotzki03}. Some artifacts remain in the
data, but they do not significantly affect the spectral analysis.

\subsection{Model atmosphere fits} 
The spectral analysis of these data was originally performed by 
\cite{Voss06}.  Since then, the models were significantly improved by
including in a consistent way the Balmer line broadening due to
simultaneous interactions with neutral and charged perturbers. An
up-to-date description of the methods and input physics is presented in
\cite{Koester09}.  We have therefore repeated the whole fitting process
for all objects; the major differences to \cite{Voss06} appear, as expected,
at the cool end of the DA sequence.

The $\chi^2$ minimization fitting routine is based on
the Levenberg-Marquardt algorithm \citep{Press92} to derive the
best fitting effective temperature and surface gravity for each
spectrum. Some more details on the fitting process can be found in
\cite{Homeier98}. For the present study we applied pure hydrogen
models and fitted the Balmer lines H$\alpha$ to H9.

To demonstrate the typical quality of the observations, we have taken
arbitrarily the entries 301 to 310 from Table~1 (only available in the
online version) and plotted the fits to the lowest six Balmer lines in
Fig.~1. The header of each panel gives the name, effective
temperature, and surface gravity of the fit.

\section{Data and atmospheric parameters for DAs} 

Table~1 lists the results of the model fitting for 615 apparently
normal DA white dwarfs.  Of those, 187 are newly identified DAs, 111
from the HES \citep[][designation HE]{Christlieb01} and 76 from the
HQS \citep[][designation HS]{Homeier98} surveys.  The primary
designation of the objects in the first column is WD, if the objects
appear in the SIMBAD or MCS databases and were known to be
spectroscopically identified white dwarfs prior to the start of the
SPY. A few objects have EC or MCT designations, if they were
identified in the Edinburgh-Cape \citep{Kilkenny91} or
Montreal-Cambridge-Tololo \citep{Demers90, Lamontagne00} surveys, but
don't seem to have WD designations.  Column 5 gives a few alternative
names. This list is not intended to be complete, but additional
information can easily be found in SIMBAD or MCS.

For the remainder of the objects we use HE or HS designations,
depending on the original catalog. In this case there are three
different meanings of column 5

\begin{itemize} 

\item if empty, there is no entry in either the SIMBAD or the MCS
database and we assume that this is a new detection of the HQS or HES.

\item if there is an entry in parentheses, this means that there is an
entry in SIMBAD, but no spectroscopic identification as a white dwarf,
at least not prior to identification in a publication related to HQS,
HES, or SPY. This mostly concerns catalogs of blue or large proper
motion objects. Some objects are also contained in more recent
catalogs, e.g. Sloan Digital Sky Survey = SDSS \citep{Adelman08}, or
Two Micron All Sky Survey = 2MASS, \citep{Cutri03}, but
not identified as white dwarfs.

\item if there is a regular entry in column 5, this describes a recent
spectroscopic identification, which was unknown at the time of our
target selection. Catalog entries here are SDSS \citep{Kleinman04,
Eisenstein06}, BGK \citep{Brown06}, Kawka06 \citep{Kawka06}.
\end{itemize}

The magnitude column lists the $V$ magnitude, if available from the SIMBAD
or MCS databases. If the number is followed by a B, this is either a Johnson B
magnitude, or, in most cases, a photographic magnitude from the MCT, HQS, or
HES surveys. mc denotes a Greenstein multichannel magnitude, obtained from
the MCS catalog. The errors of the photographic magnitudes are typically
much larger (0.1 - 0.2 mag) than suggested by the two decimal places, which
are kept only to obtain a more homogeneous table.

\begin{figure}
\includegraphics[width=8.8cm]{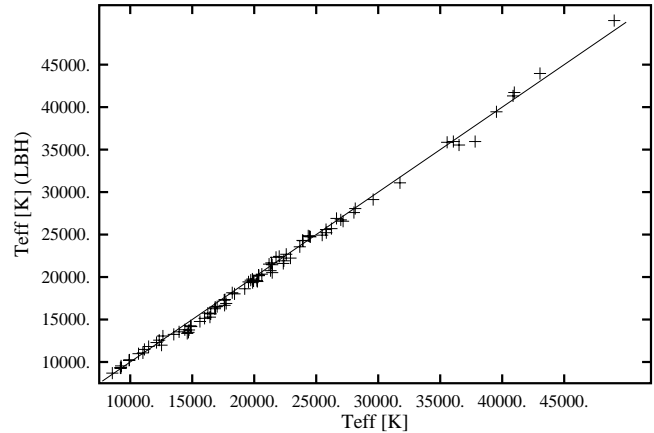}
\caption{Comparison of effective temperatures from this work with the
results of \cite{Liebert05}, (=LBH) for 85 objects in common.
  \label{fig2}}
\end{figure}

\begin{figure}
\includegraphics[width=8.8cm]{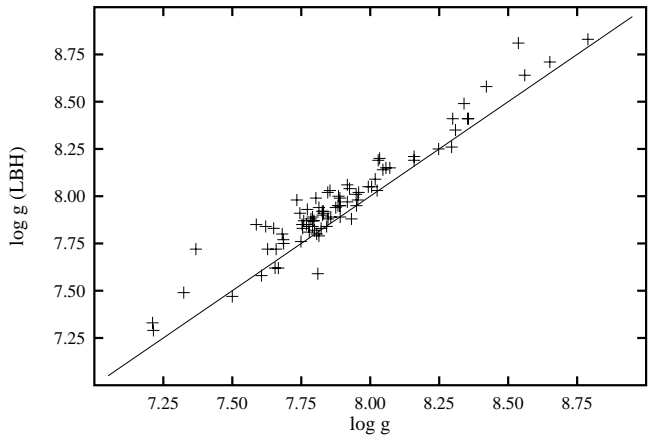}
\caption{Comparison of surface gravities from this work with the
results of \cite{Liebert05}, (=LBH) for 85 objects in common.
  \label{fig3}}
\end{figure}

Since most stars have more than one spectrum observed, the parameters
are the weighted averages of the individual solutions, with the
inverse square of the formal 1\,$\sigma$ uncertainties as weights.
The 1\,$\sigma$ final uncertainties given in Table~1 are obtained from
the individual values and should only be used as an indicator of the
quality of the data.  As is well known, with spectra of the quality
used here, the systematic errors from the reduction and fitting
process are usually much larger than the purely statistical
uncertainties. We estimated more realistic uncertainties by comparing
the differences between solutions from several spectra of the same
object. For 592 objects with multiple solutions we obtain standard
deviations of $\sigma(\Teff)\ \approx 2.5\% $ and
$\sigma(\logg)\ \approx 0.09$. These should be regarded as lower
limits, because the different spectra still used the same
observational setup, theoretical models, and fitting procedures. The
uncertainties are definitely larger than this at the high
temperature end above 50\,000~K, because NLTE effects are not
considered in our models. They are also larger, in particular for the
surface gravity, at temperatures below 8\,000~K, because the spectra
become less sensitive to this parameter, and because the neutral
broadening of the Balmer lines higher than H$\gamma$ is only
approximative.

Another method to estimate the size of the uncertainties is the
comparison of results by different authors for the same objects. The
most interesting recent study is the work by \cite{Liebert05} on the
DA white dwarfs in the Palomar Green Survey. Eliminating DAs with
\Teff\ below 8\,000~K or above 50\,000~K (see above), as well as the
double degenerates, leaves 85 objects in common. Fig.~\ref{fig2} shows
the comparison for the effective temperatures, and Fig.~\ref{fig3} for
the surface gravities. The systematic shift in \Teff\ for the whole
sample is 1.2\%, with our temperatures being slightly higher. For
\logg\ the shift is 0.08, with our values lower.  We can estimate
  the intrinsic uncertainties of our determinations by first
correcting for these systematic shifts.  Comparing the corrected
parameters with those of \cite{Liebert05}, the remaining standard
  deviations are $\sigma(\Teff) = 2.3\%$, and $\sigma(\logg) = 0.08$.
These values should be taken as indicative of the {\emph statistical}
errors of our results, and are compatible with the estimates derived
above for the internal uncertainties from multiple spectra
within our sample.

The surprisingly large systematic shift in surface gravity is 
unsatisfactory. A similar trend was also noted by \cite{Liebert05} 
in their comparison with studies by other authors. In four out of five cases
their \logg\ was higher by 0.06 - 0.10 dex. Three of these used similar
models to those used here, and the differences could arise from differences
between the "Bergeron" and the "Koester" models, or from the fitting
procedures used. We are, however, not aware of any obvious explanations
for such differences.
Systematic differences of this magnitude were also found by \cite{Napiwotzki99}
in a comparison of different studies using low-resolution spectra.

A possible reason for systematic differences may also come from the nature of
our observational data. The main purpose of the SPY was the determination
of radial velocities, which needed high resolution; therefore the echelle
spectrograph UVES was used. The wavelength interval per order ranges from 
$\approx 30$ \AA\ in the blue to $\approx 50$ \AA\ in the red region. Thus, 
e.g. H$\alpha$ at maximum strength will extend over six or more orders,
which have to be flatfielded and merged, removing the strong sensitivity
change within orders. A real flux calibration was not possible, but the
quality of the spectra obtained after some reprocessing of the ESO pipeline
results was still very impressive. Before the start of this project, we
were not certain that the spectra would be useful for anything else
except radial velocity determinations.

Nevertheless, the merging of the orders and the approximate flux calibration
attempted may have left very subtle artifacts
influencing the far wings of the strong lines, thus influencing in particular
the surface gravity results. An indication for this is visible in Table~1 in
\cite{Koester01}, which compares the results of fits to echelle vs. 
single-order low-resolution spectra for the same seven DAs. The average 
surface gravity is lower by 0.07 in the results from the echelle
spectra. Another hint towards this effect can be found in the study of
low-resolution SDSS spectra of brighter DA white dwarfs by \cite{Koester09a}, 
which used the same models and fitting routines as this work. The average
surface gravity for 578 DAs with magnitude $g < 19$, S/N $ >10$, and
8\,000 K $\le$ \Teff $\le$ 16\,000 K  is 8.014, while the value from our
current sample for the same temperature range is 7.947 from 211
objects. These are not the same objects of course, but the samples are
so large, that the difference is significant.

Also marked in Table~1 (online version) are known ZZ Ceti variables
(DAV), as well as objects, which have been observed photometrically,
but were not found to vary (NOV). The references for the NOV
designations are: (1) \cite{Gianninas05}, (2) \cite{Kepler95}, (3)
\cite{Mukadam04}, and (4) \cite{Bergeron04}.  If no reference is given
the classification is a result of the SPY and/or follow-up
observations \citep{Voss06, Voss06a, Castanheira06, Silvotti05}.  We
have not marked candidates for variability studies, but obviously all
objects in the range of \Teff\ 10\,000 - 13\,000~K are interesting in
this respect.

The Balmer lines reach their maximum strength near \Teff = 13\,000, the
exact value depending on the surface gravity. Therefore quite often two
different fitting solutions exist, which produce the same overall
strength (i.e. equivalent width) of the lines. The $\chi^2$ values of
the two minima are often similar, since model and observation
are fitted in the far line wings and differences show up only in the
inner core of the line.  Visual inspection is usually sufficient to
determine the correct solution. In a few cases, however, the
difference was so small that we preferred to give both solutions in
Table~1.

\begin{figure}
\includegraphics[width=8.8cm]{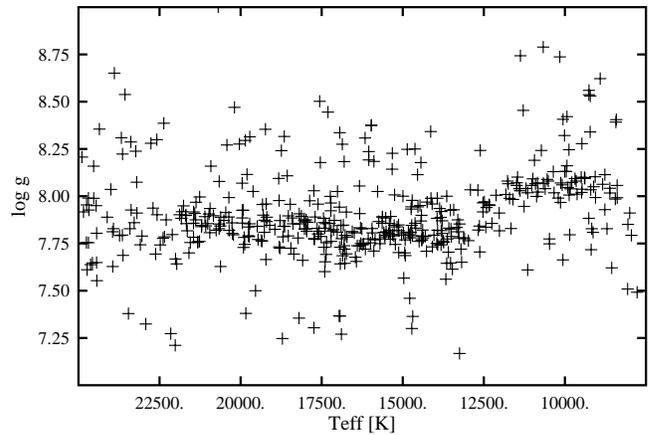}
\caption{Distribution of surface gravities for all objects between \Teff\ 
7\,500 - 25\,000~K
  \label{fig4}}
\end{figure}

Because of the selection of the targets -- as mentioned in the
introduction -- our sample is not well suited for a study of white
dwarf population characteristics such as the mass distribution.
However, it can be used to demonstrate an effect well known for many
years \citep{Bergeron90a, Bergeron92, Kleinman04, Eisenstein06,
  Kepler07, DeGennaro08, Koester09a}.  This is the fact that the
surface gravity seems to increase around \Teff $\approx$ 12\,000~K
towards lower temperatures. Figure~\ref{fig4} clearly shows this for
465 objects between 7\,500 and 25\,000~K. Taking the direct averages
without any weighting we find $<\log g >$ = 7.86 for effective
temperatures above 12\,500~K and $<\log g >$ = 8.06 below. This is
very similar to the results in the SDSS (Data Release 4) as studied by
\cite{Koester09a}. A number of possible explanations is discussed in
that study, and the most likely is found to be an inadequate
description of convection with the mixing-length
approximation. However, this problem is certainly not yet solved.

\section{Double-degenerate white dwarfs}

Two or more spectra were taken for the vast majority of the SPY
targets.  Close binaries among them could be detected with a high
level of confidence by checking for radial velocity variations
indicating orbital motion.  A total of 36 close binaries were detected
among DA sample presented here from the spectra taken for the SPY. 
This count includes only the
double degenerates, i.e.\ systems consisting of two white dwarfs.
The DA+dM systems listed in Table~3 are not included in this count. 
Seventeen of the double-degenerates are double-lined, i.e.\ spectral lines
of both white dwarfs are present in the spectra. The white dwarf companion
in the single-lined systems is already so cool and faint that it does not
produce a significant contribution to the combined spectrum.

Three more double-degenerates are marked in Table~1 as DD, but were
found by independently obtained observations: HE\,1511-0448 \citep{Nelemans05},
WD\,1241-010 \citep{Marsh95}, WD\,1022+050 and WD\,2032+188 
\citep{Morales05}.

The fitting procedure for the model atmosphere analysis is not
affected by the binary nature of the single-lined binaries. The
resulting fit parameters are those of the visible bright
component. The situation is different for the double-lined systems. In
these cases a deconvolution of simultaneous fit of both components
would be necessary for accurate parameters. We have tools for this
kind of analysis available \citep{Napiwotzki04},
but in most cases more than the two spectra
taken during the survey are needed to derive reliable parameters of
both components. Here we present the results of a fit assuming a
single star. Although these have to be taken with a pinch of salt,
they are still a useful indication of the nature of properties of the
binary. Double-lined systems are indicated in Table~1.

\section{Objects with magnetic fields or helium contamination}

\setcounter{table}{1}
\begin{table*}
\caption{Magnetic or helium-contaminated hydrogen-rich stars}
\begin{tabular}{l r r l l l}
\hline\hline
Object & RA(2000) & DEC(2000) & mag(band) & Aliases & Remarks \\ 
\hline
WD\,2359$-$434   & 00:02:10.73 & $-$43:09:55.3 & 13.05 & L362\,$-$81, LHS\,1005             & DAP\\
HS\,0051+1145    & 00:54:18.25 & +12:01:59.9   & 15.60 & (PHL\,886)                       & DAH\\
WD\,0058$-$044   & 01:01:02.25 & $-$04:11:11.2 & 15.38 & GD\,9, GR\,407                     & DAH\\
HS\,0209+0832    & 02:12:04.90 & +08:46:50.1   & 13.90 &                     & DAB\\
WD\,0239+109     & 02:42:08.54 & +11:12:31.8   & 16.18 & G\,004$-$034, LTT\,10886           & DAH\\
WD\,0257+080     & 02:59:59.24 & +08:11:55.3   & 15.90 & LHS\,5064, G\,76$-$48              & DAH\\
HS\,1031+0343    & 10:34:30.14 & +03:27:36.0   & 16.50 B&                               & DAH \\
HE\,1233$-$0519  & 12:35:37.58 & $-$05:35:36.7 & 16.48 &                                & DAH\\
WD\,1953$-$011   & 19:56:29.21 & $-$01:02:32.2 & 13.69 & G\,092$-$040, L\,0997$-$021        & DAH\\
WD\,2051$-$208   & 20:54:42.76 & $-$20:39:25.9 & 15.06 & HK\,22880$-$134                  & DAH\\
WD\,2105$-$820   & 21:13:16.52 & $-$81:49:14.3 & 13.50 & L\,24$-$52, LTT8381              & DAH\\
\end{tabular}
\end{table*}

Table~2 summarizes the data for some objects, which appear hydrogen-rich
with some peculiarities, either Zeeman splitting of the Balmer lines
due to a magnetic field or helium lines in addition to the stronger
Balmer lines. For most of the stars we obtained fits with pure
hydrogen model atmospheres. Since these are obviously not very
reliable, we do not publish the parameters here, but discuss these
stars individually.  

\begin{figure}
\includegraphics[width=8.8cm]{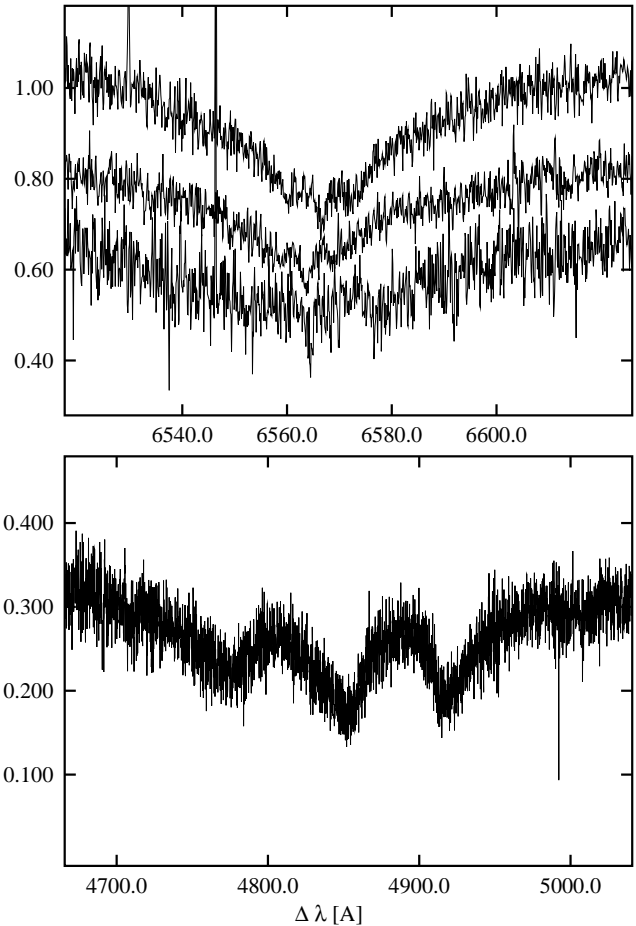}
\caption{Four new magnetic DAs. Top panel: H$\alpha$ in WD2051-208,
  HS0051+1145, HE1223-0519 (from top). Bottom: H$\beta$ in HS1031+0343.
Vertical axis is relative intensity, with arbitrary offsets between
spectra for clarity.  \label{fig5}}
\end{figure}

Four magnetic DA stars in the SPY sample have not been published
before, HS\,0051+1145, HE\,1233$-$0519, HS\,1031+0343, and
WD\,2051$-$208. Two more objects were published first as DAH stars in
\cite{Koester01}, WD\,0058$-$044 and WD\,0239+109.  Four more
magnetic DA, already described in the literature, are in the sample.
Below, these ten objects and some additional white dwarfs of special
interest are discussed.

\paragraph{HS\,1031+0343} This object is a new magnetic DA star. The
only Zeeman triplet that is completely present, and for which the
three components are well discernible, is that of H$\beta$. The
$\sigma^-$, $\pi$, $\sigma^+$ components are found at 4771\,\AA,
4851\AA, and 4905\AA. The components of the higher lines are blended
together, and the $\sigma^+$ component of H$\alpha$ is shifted by an
amount that places it outside of the observed spectral range; thus
only the $\sigma^-$ and $\pi$ components of H$\alpha$ are available in
the SPY data, at 6560\,\AA$\:$ and 6420\,\AA. The magnetic field is
estimated as $B = 6.1 \pm 0.3 $ MG.

\paragraph{WD\,0058$-$044} This star has been published as a magnetic
DA by \cite{Koester01}. The shifts of the $\sigma$ components
with respect to the $\pi$ components are $6.2 \pm 0.3$\AA$\;$ and
$3.6\pm0.3$\AA, for H$\alpha$ and H$\beta$, respectively. The
components of the higher Balmer lines are blended. The quadratic
splitting of the lines is negligible here since the split is small and
thus the field strength has to be low. The field is $B= 330
\pm 30$ kG, from H$\alpha$, and $B = 310 \pm 20$ kG, from H$\beta$.

\paragraph{WD\,0239+109} \cite{Greenstein90} recognized an
unusual line shape for this object, and suggested a magnetic field as
one of the possible reasons. \cite{Bergeron90b} interpreted the
spectrum as that of an unresolved DA+DC binary. The SPY spectra in
\cite{Koester01} revealed the presence of Zeeman splitting of
H$\alpha$ and thus proved the magnetic nature of the object. In the
first SPY spectrum, the H$\alpha$ components are placed at
6576.6\,\AA, 6562.3\,\AA, and 6548.4\,\AA, and those of H$\beta$ at
4868.9\,\AA, 4861.0\,\AA, 4853.0\,\AA, and from that field strengths
of $700 \pm 20$ kG, from H$\alpha$, and $720 \pm 30$ kG, from
H$\beta$, can be derived.

\paragraph{WD\,0257+080} \cite{Bergeron97} found a flat-bottom
H$\alpha$ core for this object which is typical for white dwarfs with
a low-strength magnetic field, but they were not able to identify a
Zeeman triplet and estimated the field strength to be $\sim 100$ kG.
One of the two SPY spectra of WD\,0257+080 clearly shows an H$\alpha$
triplet.  In the other spectrum, which has a slightly lower $S/N$,
only the $\sigma^-$ component is discernible. The H$\alpha$ components
of the first spectrum are split by $ 1.8 \pm 0.3\,$\AA, from which
$B=90 \pm 15$ \,kG can be derived, a more precise value than the
previous estimate.

\begin{table*}
\caption{Data for DA+dM binaries. In boldface are the new SPY detections. 
For explanations of the magnitude column see text.}
\label{dadmtable}      %
\begin{tabular}{l r r l l l}        
\hline\hline                 
Object & RA(2000) & DEC(2000) &  mag(band) & Aliases & Type \\    
\hline                        
{\bf HE\,0016$-$4340} & 00:19:06.10 & $-$43:24:18.5 & 15.62 B &                                    & DA2\\
WD\,0034$-$211        & 00:37:24.99 & $-$20:53:43.6 & 14.53   & MCT\,0034$-$2110, LP\,852$-$559    & DA3\\
{\bf HE\,0105$-$0232} & 01:08:06.33 & $-$02:16:53.1 & 15.77 B & (PB6272)                           & DA2\\ 
WD\,0131$-$163        & 01:34:24.08 & $-$16:07:08.2 & 13.98   & MCT\,0131$-$1622, PHL\,1043        & DA1\\
WD\,0137$-$349        & 01:39:42.88 & $-$34:42:39.1 & 15.33   & HK\,29504$-$36                     & DA3\\
WD\,0205+133          & 02:08:03.59 & +13:36:23.9   & 13.78 B & PG\,0205+134                       & DA1\\ 
WD\,0232+035          & 02:35:07.67 & +03:43:55.6   & 12.25   & Feige\,24                          & DA1\\
WD\,0303$-$007        & 03:06:07.21 & $-$00:31:14.3 & 16.21   & HE\,0303$-$0042, KUV\,03036$-$0043 & DA2\\ 
WD\,0308+096          & 03:10:54.90 & +09:49:31.6   & 15.23   & PG\,0308+096                       & DA2\\ 
HE\,0331$-$3541       & 03:33:52.53 & $-$35:31:18.9 & 14.79 B &                                    & DA2\\ 
WD\,0347$-$137        & 03:50:14.55 & $-$13:35:13.5 & 14.00   & GD\,51, LP\,713$-$034              & DA3\\ 
{\bf HE\,0409$-$3233} & 04:11:21.14 & $-$32:26:14.9 & 16.03 B &                                    & DA3\\
WD\,0429+176          & 04:32:23.76 & +17:45:02.4   & 13.93   & GH\,7$-$255, HZ9B                  & DA3\\ 
WD\,0430+136          & 04:33:10.59 & +13:45:12.4   & 16.50   & KUV\,04304+1339                    & DA1\\ 
{\bf HE\,0523$-$3856} & 05:25:28.11 & $-$38:54:12.5 & 16.07 B &                                    & DA3\\ 
WD\,0718$-$316        & 07:20:47.92 & $-$31:47:04.6 & 15.10   & RE\,0720$-$314, EUVE\,J\,0720$-$317& DAO\\
WD\,0933+025          & 09:35:40.69 & +02:21:59.6   & 16.01 B & PG\,0933+026                       & DA2\\ 
WD\,0950+185          & 09:52:45.80 & +18:21:02.9   & 15.30   & PG\,0950+186                       & DA2\\ 
WD\,1001+203          & 10:04:04.30 & +20:09:22.5   & 15.35   & TON\,1150, HS1001+2023             & DA2\\ 
WD\,1026+002          & 10:28:34.88 & $-$00:00:29.6 & 13.89 B & PG\,1026+002, HE\,1026+014         & DA3\\ 
WD\,1042$-$690        & 10:44:10.63 & $-$69:18:22.9 & 13.09   & BPM\,06502, L\,101$-$080           & DA2\\ 
WD\,1049+103          & 10:52:27.82 & +10:03:36.5   & 15.83 B & PG\,1049+103                       & DA3\\ 
HE\,1103$-$0049       & 11:06:27.66 & $-$01:05:14.9 & 16.30 B &                                    & DA3\\ 
{\bf HE\,1208$-$0736} & 12:11:01.08 & $-$07:52:42.9 & 15.67 B &                                    & DA2\\ 
WD\,1247$-$176        & 12:50:22.13 & $-$17:54:48.2 & 16.19   & EC\,12477$-$1738, HE\,1247$-$1738  & DA3\\ 
WD\,1319$-$288        & 13:22:40.46 & $-$29:05:35.0 & 15.99   & EC\,13198$-$2849                   & DA5\\ 
HE\,1333$-$0622       & 13:36:19.64 & $-$06:37:58.9 & 16.05   & WD\,1333$-$063                     & DA2\\ 
WD\,1334$-$326        & 13:37:50.77 & $-$32:52:22.5 & 16.34   & EC\,13349$-$323                    & DA1\\
{\bf HE\,1346$-$0632} & 13:48:48.34 & $-$06:47:21.0 & 16.27 B &                                    & DA2\\ 
EC\,13471$-$125       & 13:49:51.95 & $-$13:13:37.5 & 14.80   & WD\,1347$-$129, RXS                & DA3\\
WD\,1415+132          & 14:17:40.22 & +13:01:48.6   & 15.29   & US\,3974, Feige\,93                & DA1\\ 
EC\,14329$-$162       & 14:35:45.70 & $-$16:38:17.0 & 14.89   & WD\,1432$-$164, HE\,1432$-$1625    & DA3\\
WD\,1436$-$216        & 14:39:12.70 & $-$21:50:14.6 & 15.94   & EC\,14363$-$2137, HE\,1436$-$2137  & DA2\\ 
WD\,1458+171          & 15:00:19.36 & +16:59:14.7   & 16.12 B & PG\,1458+172                       & DA5\\ 
WD\,1541$-$381        & 15:45:10.97 & $-$38:18:51.3 & 14.90   & LDS\,539B, L\,480$-$085            & DA4\\ 
{\bf HS\,1606+0153}   & 16:08:55.22 & +01:45:48.6   & 15.00 B &                                    & DA3\\ 
WD\,1643+143          & 16:45:39.05 & +14:17:42.0   & 15.38   & PG\,1643+144                       & DA2\\ 
WD\,1646+062          & 16:49:07.83 & +06:08:43.6   & 15.84 B & PG\,1646+062                       & DA2\\ 
WD\,1845+019          & 18:47:39.09 & +01:57:33.5   & 12.95   & LAN\,18, KPD\,1845+0154            & DA2\\
WD\,1844$-$654        & 18:49:02.00 & $-$65:25:14.2 & 15.80 B & HK\,22959$-$81                     & DA1\\
{\bf HS\,2120+0356}   & 21:23:09.53 & +04:09:28.3   & 16.20 B &                                    & DA3\\
{\bf HE\,2123$-$4446} & 21:26:41.88 & $-$44:33:38.8 & 15.97 B &                                    & DA3\\ 
{\bf HE\,2147$-$1405} & 21:50:03.69 & $-$13:51:45.9 & 15.94 B & (PHL\,167)                         & DA2\\ 
WD\,2151$-$015        & 21:54:06.53 & $-$01:17:10.9 & 14.41   & G\,093$-$053,L\,1003$-$016         & DA2\\
{\bf HE\,2217$-$0433} & 22:20:05.80 & $-$04:18:44.5 & 16.28 B &                                    & DA3\\ 
WD\,2313$-$330        & 23:16:02.90 & $-$32:46:41.4 & 15.74 B & HK\,22888$-$45, MCT                & DA1\\
\hline                           
\end{tabular}
\end{table*}

\paragraph{WD\,1953$-$011} A weak Zeeman-split triplet of H$\alpha$
was found by \cite{Koester98}, from which they derived a field
strength of 93\,kG. \cite{Maxted00} had noticed a variable
depression in the wings of H$\alpha$ which they identified as
additional Zeeman-split H$\alpha$ features, which led them to assume a
non-simple field geometry with a strong spot-like field of $\sim
500\,$kG combined with a weaker 70\,kG dipole field.  Comparing the
two SPY spectra, the H$\alpha$ wings appear deformed near 6550\,\AA\
and 6575\,\AA, which might allow a very rough estimate of field
strengths of $\sim$ 500\,kG up to $\sim$ 750\,kG, if it is assumed
that these features are of magnetic origin. However no variation of
these features as described by Maxted et al. can be found, the line shape is
very similar in both SPY spectra. The central triplet however is
obvious, and the splitting of the core is $1.9 \pm 0.2\,$\AA. This
corresponds to a dipole $B$ field of $95 \pm10 \,$kG.

\paragraph{WD\,2105$-$820} The spectrum of this star was found to show
a flat-bottom H$\alpha$ core, probably due to a low magnetic field, by
Koester et al. (1998), from which they derive a field strength of
43\,kG. No Zeeman triplet is obvious in the SPY spectra as well, they
also only exhibit the broadened, flat core; the full width of the core
of 1.8\,\AA$\:$ in the SPY spectra is consistent with the field
strength derived by \cite{Koester98}.

\paragraph{WD\,2359$-$434} \cite{Koester98} suspected that this
DA could be magnetic due to its flat H$\alpha$ core, and a very low
field of only 3\,kG was polarimetrically found by \cite{Cuardrado04}.
 They selected this object as one of their program stars
based on the criterion that the SPY spectra show no signs of Zeeman
splitting; however they themselves note that flat Balmer line cores
are present in the spectra of that object, and if these are caused by
the $B$ field, it would indicate a higher field strength than that
derived from the polarimetry data. \cite{Kawka07} measure a low field
of $3.4 \pm 4.4$ kG.

The SPY spectra indeed do not only show a flat-bottom H$\alpha$ core,
but within that core also a pronounced $\pi$ component and less clear,
broad $\sigma$ components, centered at 6561.1\,\AA, 6563.5\,\AA, and
6565.8\,\AA. Thus a field strength of $110 \pm 10$ kG can be
derived. This is two orders of magnitude stronger than
found by \cite{Cuardrado04} and \cite{Kawka07}.  The reason for
these different results is unclear.

\paragraph{WD\,0446$-$789 and WD\,1105$-$048} These objects are the
two remaining of the three for which \cite{Cuardrado04} discovered
magnetic fields of only a few kG from polarimetry data.  The H$\alpha$
core of WD\,0446$-$789 appears slightly broadened, the width is
1.8\,\AA, which is slightly wider than the average line core width of
$\sim 1$ \AA. If we interpret this excess width as due to magnetic
broadening, this would correspond to a field strength on the order of
10\,kG.  The line core of WD1105-048 shows no peculiarities, it has a
normal width of 1\,\AA\ and thus no detectable field.

\paragraph{HS\,0051+1145} This object has previously been found as a blue
source, PHL\,886, but was not observed spectroscopically before
the SPY. It is a new magnetic DA. The SPY spectra have a rather low $S/N$
and thus only the H$\alpha$ triplet of one of the spectra is
resolved. The components are placed at 6558.9\,\AA, 6563.6\,\AA, and
6568.6\,\AA, yielding an approximate field strength of $240 \pm 10$
kG.

\paragraph{HE\,1233$-$0519} This DA was published by \cite{Koester01},
but not recognized as a magnetic star. The SPY spectra
have a low $S/N$ such that only the H$\alpha$ triplet is discernible
in one of the spectra. With the components at 6552.2\,\AA,
6564.4\,\AA, and 6576.6\,\AA, a field strength of $610 \pm 10$ kG
results.

\paragraph{WD\,2051$-$208} \cite{Beers92} published this object
as a DA, but it was not further investigated since. It is a new
magnetic DA, and shows a variable Zeeman splitting of H$\alpha$ and
H$\beta$. The H$\alpha$ components in the two SPY spectra are found at
6562.0\,\AA, 6566.6\,\AA, 6570.7\,\AA, and at 6560.2\,\AA,
6566.6\,\AA, 6571.9\,\AA, respectively, and those of H$\beta$ at
4861.2\,\AA, 4864.0\,\AA, 4866.7\,\AA, and at 4861.0\,\AA,
4863.9\,\AA, 4866.9\,\AA. The resulting field strengths are $220 \pm
20$ kG and $290 \pm 20$ kG (from H$\alpha$) as well as $250 \pm
30$ kG and $270 \pm 30$ kG (from H$\beta$). The values derived
from H$\alpha$ are significantly different, and each is consistent
with the corresponding value from H$\beta$.

\paragraph{WD\,2253$-$081 and WD\,1344+106} \cite{Bergeron01}
suspected that shallow line cores which they found for these objects
might indicate low-strength magnetic fields. The line cores of both
objects have since been fitted with rotationally broadened line
profiles, corresponding to projected rotation velocities of
$36^{+14}_{-7}\,$km\,s$^{-1}$ for WD\,2253$-$081 \citep{Karl05}
and $4.5\pm2\,$km\,s$^{-1}$ for WD\,1344+106 \citep{Berger05}.
The H$\alpha$ line cores in the SPY spectra of both objects neither
show Zeeman triplets nor flat bottoms that could indicate the presence
of a $B$ field. They are non-magnetic objects.

\paragraph{HS0209+0832} This DAB star is well studied since it is one
of very few objects that exhibit helium features at a temperature that
places it in the DB gap \citep{Jordan93}. \cite{Heber97} found a
helium abundance that is variable at timescales of a few months, which
has been interpreted as a sign of helium accretion from a clumpy
interstellar medium. The equivalent widths of the \ion{He}{i} 4471\,\AA$\:$
and 5876\,\AA$\:$ lines show no significant differences between both
SPY spectra, i.e., no variation of the abundance is found. This is
however not surprising since the spectra were recorded within 3 days
of each other.

Zeeman splitted H$\alpha$ or H$\beta$ for the four new magnetic DAs
are displayed in Fig.~\ref{fig5}.

\section{Binaries with DA white dwarfs and dM companions} Table~3
gives the data on binaries containing a DA and a M dwarf companion,
identified from molecular features in the red spectrum and/or Balmer
emission components. The names in boldface indicate new detections
from the SPY. The magnitude is the Johnson $V$ magnitude, unless
indicated otherwise (see description for Table~1).  The type is
estimated from the effective temperature obtained with a fit with
hydrogen models, using only the higher Balmer lines from H$\gamma$.

\section{Conclusions} 
We present the data -- coordinates, magnitudes, and alias names -- for
the hydrogen-rich objects in the SPY sample. These include 615 objects
with pure hydrogen spectra, for which atmospheric parameters derived
from fits with hydrogen models are given in Table~1. Of these, 187 are new white
dwarf detections from this survey, or the HES and HQS surveys used to
define the target list. In addition to the 615 DAs, our sample also
includes 46 DA+dM binaries, of which 10 are new, and 10 magnetic DA (4
new).  The results show that with careful reduction even
high-resolution echelle spectra can be used to determine stellar
parameters through line profile fitting, although the line profiles
may extend over many echelle orders. However, there is an indication
that the surface gravities obtained are lower by 0.05-0.08 dex,
compared to results from high S/N low-resolution spectra. The surface
gravities of the normal DAs show the well known, but still
unexplained, trend to a larger value (by 0.2 dex) for temperatures
below approximately 12500~K.

\acknowledgement{T.L. is supported within the framework of the
  Excellence Initiative by the German Research Foundation (DFG)
  through the Heidelberg Graduate School of Fundamental Physics (grant
  number GSC 129/1). Research at Bamberg in the context of the SPY
  project was funded by the DFG through grants Na 365/2-1/2 and He
  1356/40-3/4.
  This research has made extensive use of the
  SIMBAD database, operated at CDS, Strasbourg, France. }

\Online \begin{appendix} 
\section{Fit results for DA white dwarfs}
Data and fit results for mostly normal DA white dwarfs are given in
Table~1, which also includes some double-degenerate binaries
(DD). These were fitted with normal DA atmosphere models, and the
parameters should be used with some care. See the main text for
explanations of the names and aliases used in columns 1 and 5,
and their relation to the source of the first white dwarf
spectroscopic identification, as well a for the magnitudes in column 4.

In a few cases we list two different solutions for \Teff\ and \logg\
(remark amb for ambiguous), because even with close visual inspection
of the fit we were not able to select the hotter or cooler
solution. This may sometimes be helpful to understand, why a DA
apparently outside the ZZ Ceti instability strip is pulsating or vice
versa.

Meaning of the remark column:\\
DDs: double-degenerate binary with one line system\\
DDd: double-degenerte binary with lines from both companions\\
DAV:  known ZZ Ceti variable, references can be found e.g. in the Simbad database\\
NOV: observed as ZZ Ceti candidate but not found to vary\\
amb: two solutions possible, which could not be distinguished\\

The references for the NOV are: (1) \cite{Gianninas05}, (2) \cite{Kepler95},
 and (3) \cite{Mukadam04}, and (4) \cite{Bergeron04}.
If no reference is given the classification is a result of the SPY and/or
follow-up observations \citep{Voss06, Voss06a, Silvotti05, Castanheira06}.

\onllongtabL{1}{
\begin{landscape}
\begin{longtable}{l r r l l r r r r r l}
\caption{Fit results for hydrogen atmosphere stars}\\
\hline\hline
Object & RA(2000) & DE(2000) & mag(band) & Aliases & $T_{\mathrm{eff}}$ [K]& $\sigma(T_{\mathrm{eff}})$ [K] & $\log g$ &
$\sigma(\log g)$ & Spectra & Remarks \\ 
\hline  
\endfirsthead
\caption{Continued.} \\
\hline
Object & RA(2000) & DE(2000) & mag(band) & Aliases & $T_{\mathrm{eff}}$ [K]& $\sigma(T_{\mathrm{eff}})$ [K] & $\log g$ &
$\sigma(\log g)$ & No. of Spectra & Remarks \\ 
\hline  
\endhead
\hline
WD\,2359$-$324   & 00:02:32.36 & $-$32:11:50.7 & 15.87 B & MCT\,2359$-$3228                 & 22478 &   57 & 7.742 &  0.008 &   3 & \\
WD\,0000$-$186   & 00:03:11.21 & $-$18:21:57.6 & 16.29 B & MCT\,0000$-$1838, GD\,575          & 15264 &   43 & 7.846 &  0.009 &   2 & \\
HS\,0002+1635    & 00:04:43.34 & +16:52:16.7   & 15.30 B & (PHL\,647)                       & 25878 &   36 & 7.859 &  0.005 &   2 & \\
WD\,0005$-$163   & 00:07:34.80 & $-$16:05:31.8 & 16.29 mc& G\,158$-$132, GR\,509              & 15141 &   25 & 7.594 &  0.009 &   2 & NOV(1)\\
                 &             &               &         &                                & 11684 &   31 & 7.689 &  0.012 &   2 & amb \\
WD\,0011+000     & 00:13:39.19 & +00:19:23.1   & 15.31   & G\,031$-$035, WOLF\,1, SDSS        &  9472 &    7 & 8.037 &  0.010 &   2 & \\
WD\,0013$-$241   & 00:16:12.63 & $-$23:50:06.4 & 15.36   & MCT\,0013$-$2406, GR\,458          & 18529 &   26 & 7.896 &  0.005 &   2 & \\
WD\,0016$-$258   & 00:18:44.49 & $-$25:36:42.2 & 16.07 B & MCT\,0016$-$2553                 & 10828 &   14 & 8.022 &  0.008 &   2 & DAV\\
\end{longtable}
\end{landscape}
}

\end{appendix}

\end{document}